\documentclass[prl,aps,showpacs]{revtex4}
\usepackage{graphicx} 

\begin{document}
\title{Crystallization of the Kob-Andersen binary Lennard-Jones liquid}
\author{S{\o}ren Toxvaerd, Thomas B. Schr{\o}der, and Jeppe C. Dyre}
\affiliation{DNRF centre  ``Glass and Time,'' IMFUFA, Department of Sciences, Roskilde University, Postbox 260, DK-4000 Roskilde, Denmark}
\date{\today}

\begin{abstract}
At density $\rho=1.2$ the standard Kob-Andersen binary Lennard-Jones liquid crystallizes in the temperature interval $[0.39, 0.44]$ after molecular dynamics simulations of typically 10-100 $\mu$s in Argon units. The crystallization is associated with a phase separation where the large (A) particles cluster in a volume void of B particles. We investigate a modification of the  Kob-Andersen system where the attraction between particles of the same type is removed, thus disfavoring phase separation. We have not been able to crystallize this new system. 
\end{abstract}


\maketitle

As computers get faster, simulations of the highly viscous liquid phase preceding glass formation become increasingly realistic. In this context it is nice to have a standard model system to refer to, just like the Ising model is a standard model for critical phenomena. For several years the Kob-Andersen 80-20 binary Lennard-Jones (BLJ) mixture \cite{kob} has served this purpose, because it is easy to simulate and was never found to crystallize. The Kob-Andersen BLJ consists of two types of Lennard-Jones particles, 80\% large particles (A) and 20\% small particles (B). The BLJ potentials are modifications of the potentials devised by Weber and Stillinger \cite{web}, who constructed the pair potentials for the binary mixture based of physical-chemical data for the Ni$_{80}$P$_{20}$ alloy. The Kob-Andersen potentials do not obey the Lorentz-Berthelot rules for mixture, especially with respect to the A-B attraction that is three times stronger than the B-B attraction. This ensures a non-ideal mixture with large negative mixing enthalpy (and energy), which suppresses phase separation.

\begin{figure}\begin{center}
\includegraphics[height=7cm]{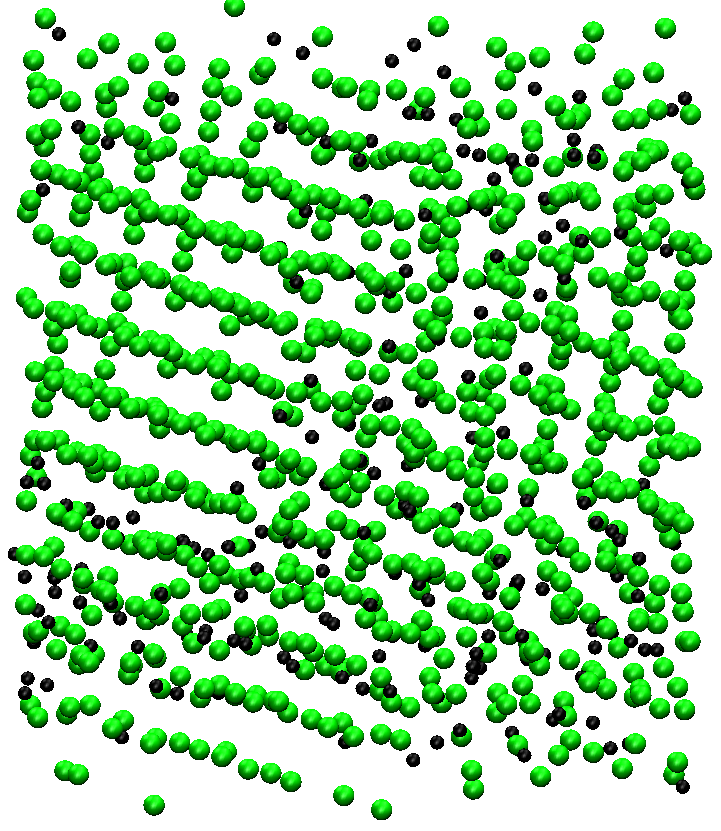}
\includegraphics[height=7cm]{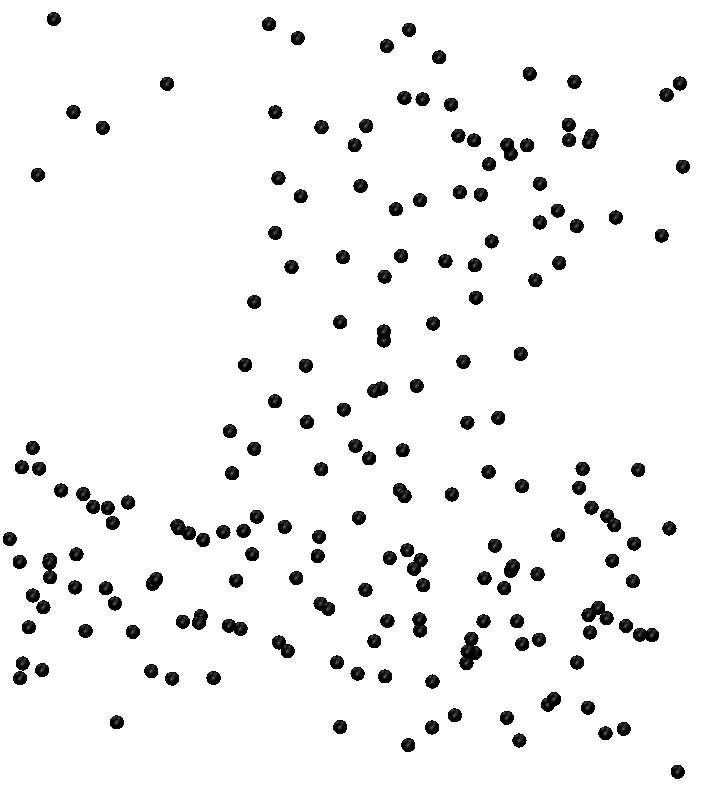}
\caption{(a) Particle positions projected onto a plane for the standard Kob-Andersen binary Lennard-Jones liquid (the large (A) particles are green, the small (B) particles are black). The system is shown after 15 $\mu$s of simulation (Argon units) at T=0.40 and density 1.2 (dimensionless units). After this simulation time the A particles phase separated and formed a large crystal, as is clear in the second figure showing the same configuration with all A particles removed.}
\end{center}\end{figure}

Recently, Pedersen and co-workers reported crystallization of the Wahnstrom BLJ \cite{wah91} after lengthy computer runs \cite{ped07}. The crystal is the ${\rm MgZn_2}$ phase consisting of particles in the ratio 1:2, different from the 1:1 ratio defining the Wahnstrom BLJ. Thus concentration fluctuations precede (or at least correlate with) crystal formation. These findings inspired us to investigate whether the more widely studied Kob-Andersen BLJ also crystallizes in lengthy computer runs.

We performed molecular dynamics simulations of 1000 particles of the standard Kob-Andersen binary Lennard-Jones liquid in the NVE and NVT ensembles (in the latter case using the standard Nose-Hoover thermostat). The system was simulated at the density 1.2 in dimensionless units \cite{kob} at varying temperatures. The standard time-reversible leap-frog algorithm \cite{tox1} was used with a time step of 0.005 (in Argon units: $\approx 10^{-14}$ s). The software used has been described elsewhere \cite{toxsoft}; it utilizes a double sorting of neighbor particles that makes it possible to simulate 10 $\mu$s within 2-3 days of computing on a standard computer.

For temperatures above 0.45 no crystallization was detected. In the temperature interval $[0.39,0.45]$ a suspicious drop in pressure taking place typically after a few billion time steps indicates that something is happening. It turns out that the system phase separates such that the A particles form fairly large regions with no B particles present. Linked to this phase separation is a crystallization of the A particles. Ten simulations were performed in the $[0.39,0.45]$ temperature interval; after 4 billion time steps eight of these ten simulations phase separated with crystallization of the A particles. Figure 1 shows a representative example of our results, giving the positions of the particles after a simulation time of 15 $\mu$s and for T=0.40 (NVT-MD). There is a large region of pure A particles showing clear crystalline order. This is consistent with the results of Fernandez and Harrowell \cite{far03}, who found that the $T=0$ equilibrium phase of the Kob-Andersen BLJ consists of a coexisting A (fcc) and AB (CsCl structure) crystals.

We proceed now to make a modification of the Kob-Andersen system, with the goal to make it less prone to crystallization. A liquid mixture is stable against crystallization if the composition of the mixture disfavors the creation of a critical nucleus of the crystal when the liquid is cooled down. In the standard Kob and Andersen (KA) system this is achieved by letting the smaller B-particles act as ``glue'' between the A-particles and thus the LJ-parameters: $\sigma_{\textrm{AB}}=0.8 \sigma_{\textrm{AA}}$; $\sigma_{\textrm{BB}}=0.88 \sigma_{\textrm{AA}}$; $\epsilon_{\textrm{AB}}=1.5 \epsilon_{\textrm{AA}}$ and $\epsilon_{\textrm{BB}}=0.5 \epsilon_{\textrm{AA}}$ disobeys the Lorentz-Berthelot (LB) mixing rules
\begin{eqnarray}
 \sigma_{\textrm{AB}}=(\sigma_{\textrm{AA}}+\sigma_{\textrm{BB}})/2,\\
 \epsilon_{\textrm{AB}}=(\epsilon_{\textrm{AA}}\epsilon_{\textrm{BB}})^{1/2}.
\end{eqnarray}

As shown above, the glue effect between A and B particles is not strong enough; the A particles eventually phase-separate and crystallize. Our modification of the KA system increases the relative strength of the attraction between A and B particles, simply by completely removing the attraction between particles of the same type, thus disfavoring phase-separation. This is done by cutting the potential between  particles of the same type in the minimum:
\begin{eqnarray}
  &u_{ij}(r_{ij})= \left \{ \begin{array}{ll}
  4\epsilon_{ij}\left[\left(\frac{\sigma_{ij}}{r_{ij}}\right)^{12}-
  \left(\frac{\sigma_{ij}}{r_{ij}}\right)^6\right]-u_{\textrm{LJ}}(r_{ij}(cut)), &
  r_{ij}< r_{ij}(cut) \\
  0, & r \geq  r_{ij}(cut), \\
\end{array} \right.
\end{eqnarray}
with
\begin{eqnarray}
  r_{i,i}(\textrm{cut})=2^{1/6}\sigma_{i,i}\\
  r_{\textrm{A,B}}(\textrm{cut})=2.5 \sigma_{\textrm{A,B}}
\end{eqnarray}
 
With this modification we have effectively prevented phase-separation with a pure A-phase, and thereby prevented the type of crystallization reported above for the standard KA system. Obviously, there is still the possibility of crystallization into more complex crystals as observed eg. in the Wahnstrom system \cite{ped07}. Indeed at the composition $x_B=0.5$ we find that the system quickly crystallizes into a CsCl crystalline structure. However at the  composition $x_B=0.2$ that is our focus here, none of the lengthy simulation runs reported below resulted in crystallization. 

A secondary advantage of our modified system is that the short cut-offs makes it much faster to simulate by using a two-step sorting of nearest neighbors \cite{toxsoft}. In additional to that one can  use a bigger value of the time increment in the MD simulation due to lower temperatures (see below), so not only does the mixture not crystallizes; but it can also be cooled in equilibrium to lower temperatures and followed over much longer times ($ms$), see Figs. 3 and 4 below.

\begin{figure}
\includegraphics[width=8.5cm]{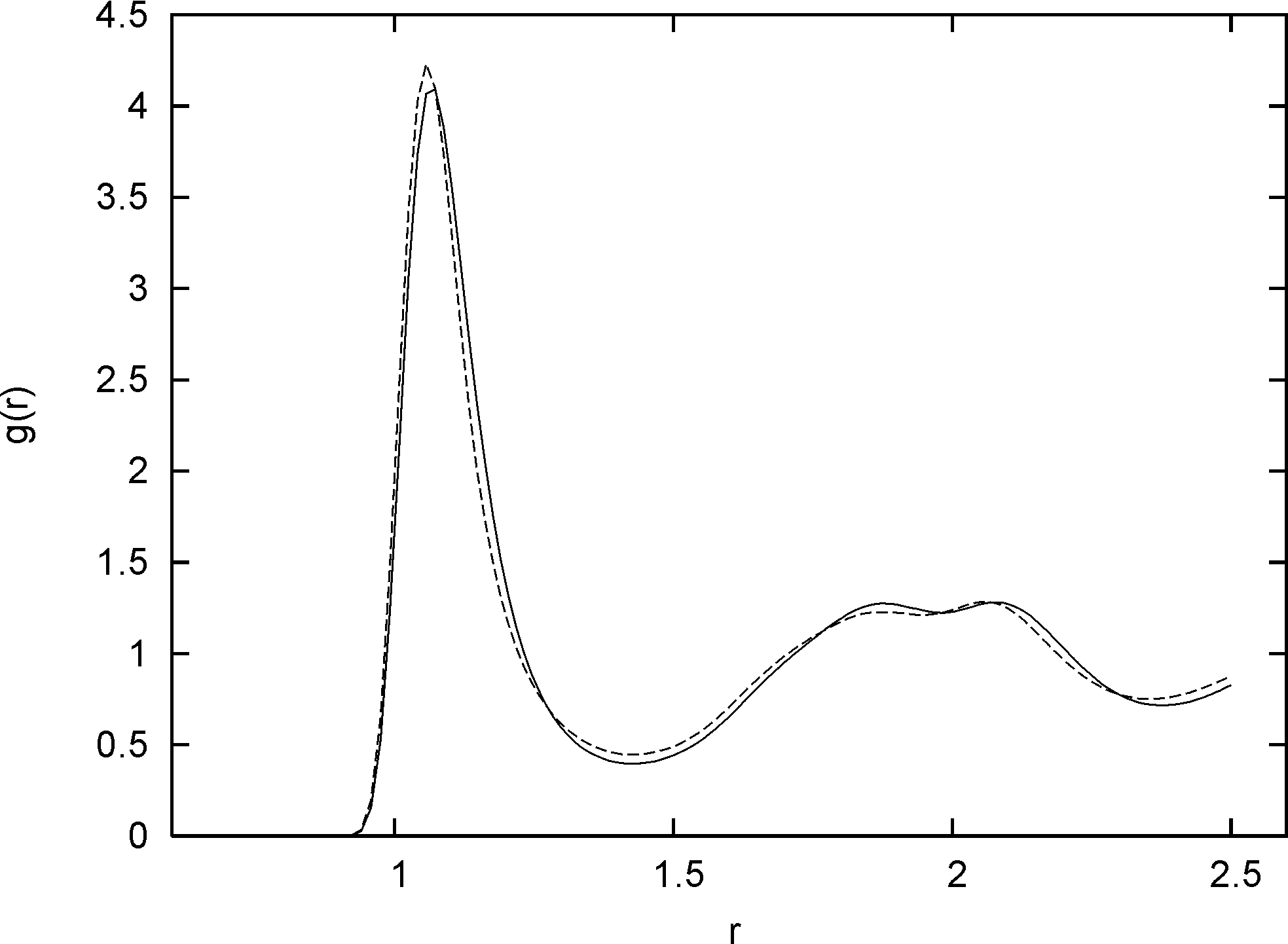}
\includegraphics[width=8.5cm]{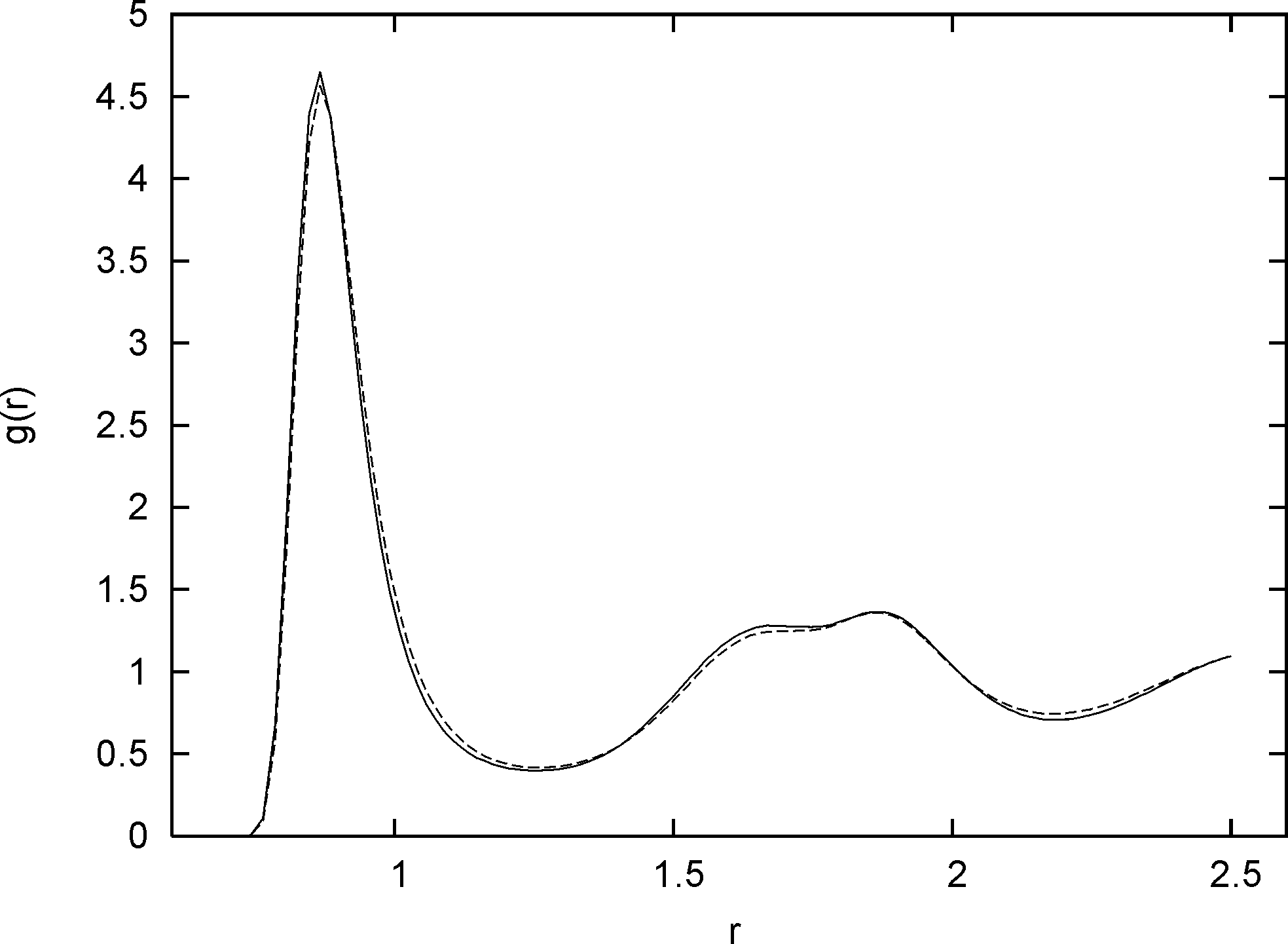}  
\includegraphics[width=8.5cm]{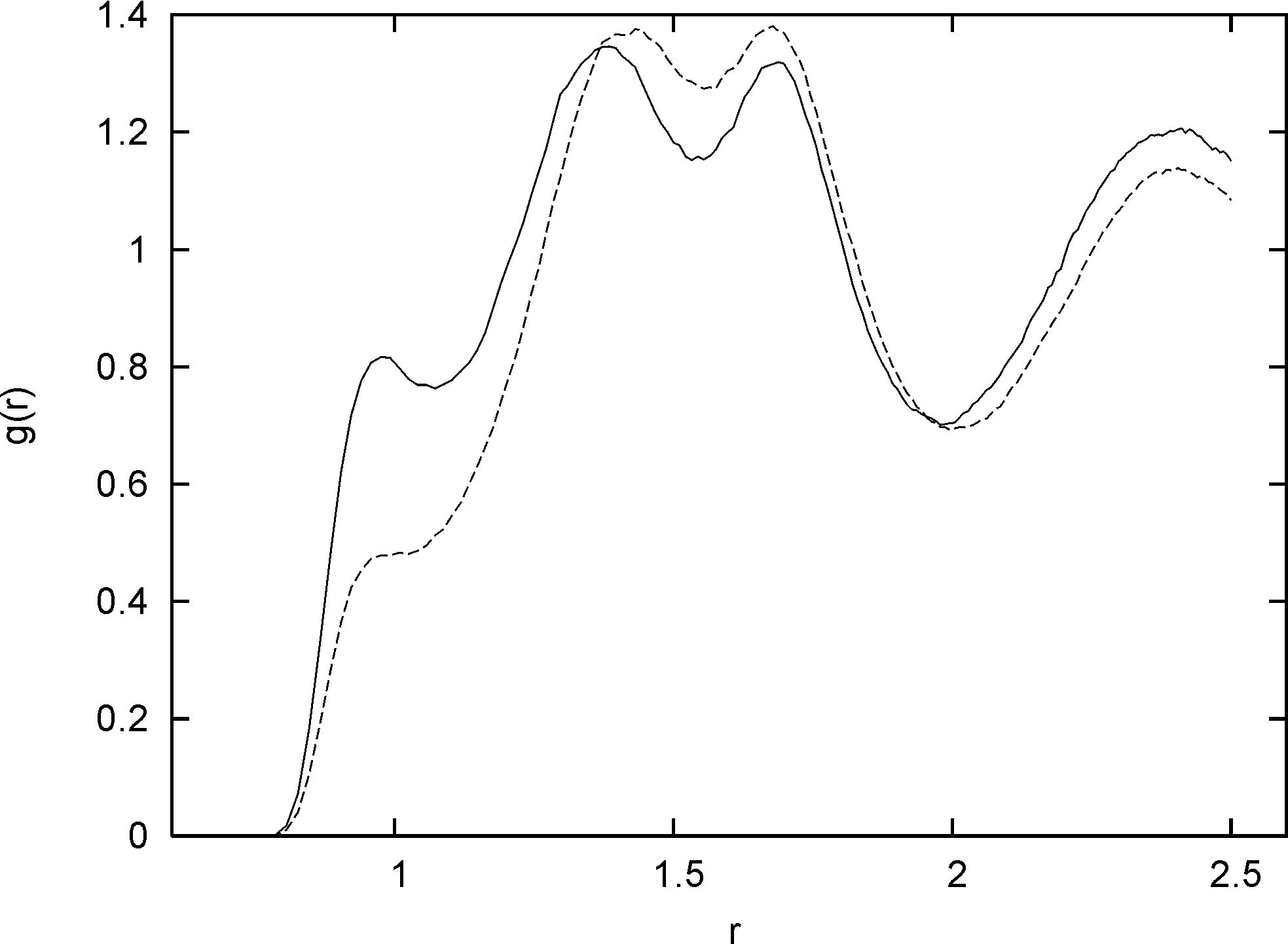}
  \caption{ Radial distribution functions $g_{\alpha,\beta}(r)$ at the density $\rho$=1.2 and the temperature T=0.50 for mixtures with particle fraction $x_{\textrm{B}}$=0.2 of B-particles. Full line is for a the standard KA-mixture and the dashed curve is for our modified mixture. (a) $g_{\textrm{A,A}}(r)$ (b) $g_{\textrm{A,B}}(r)$ (c) $g_{\textrm{B,B}}(r)$} 
\end{figure}

The structure of the modified mixture is almost the same as in the standard KA-mixture (with a particle fraction, $x_{\textrm{B}}$=0.2). Figure 2 show the radial distribution functions, $g_{\alpha,\beta}(r)$ for the modified mixture and for the KA-mixture. The similarity is consistent with what expected from the WCA-approach \cite{wca}, where the repulsive LJ forces has been used in perturbation theories for dense liquids. As can be seen from the figure 2c the distribution of the B-particles \emph{is} slightly affected. This change in  $g_{\textrm{B,B}}(r)$ is due to the relative increased "binding-energy" between the solvent A-particles and the solute of B-particles which, however, is not so visible in  $g_{\textrm{A,B}}(r)$ (Fig. 2b) due to the many A-B neighbors (notice the different y-scales of the two figures).

\begin{figure}
 \includegraphics[width=12cm]{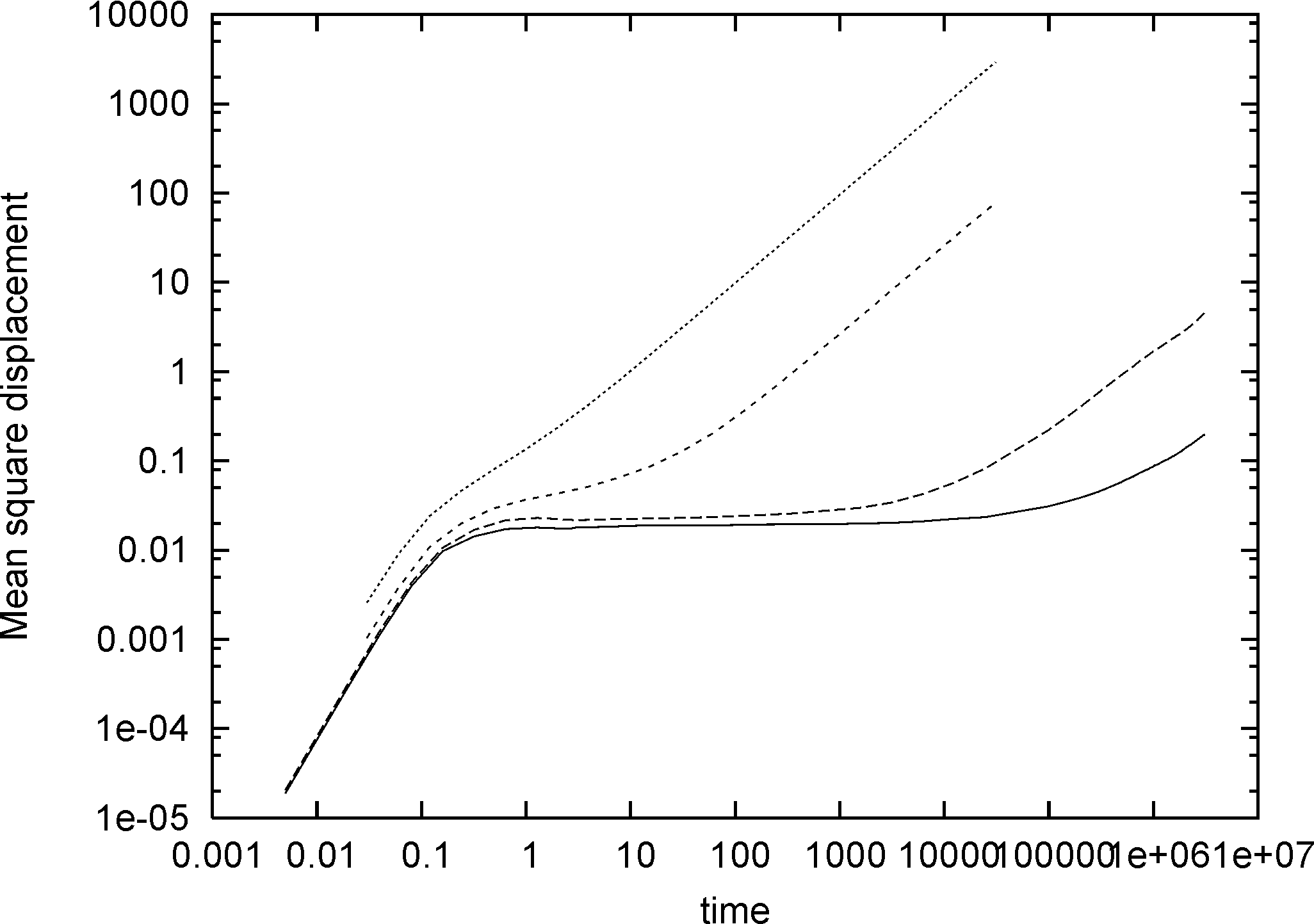}
  \caption{Log-log plot of the mean square displacement for A-particles as a function of time and
   for four different temperatures: T=1.00 with dots, T=0.40 with small dashes,  T=0.275 with long dashes and T=0.25 with full line}
\end{figure}

The KA- and the modified mixture were cooled down in steps from T=1.5. As described above, the standard KA- mixture crystallized after long runs  in the temperature interval T $\in [0.39,0.45]$; but it is possible to cool the system down to T=0.375 and still determine the self diffusion constant, D. The modified mixture was cooled down to T=0.25, where the system was followed $10^{11}$ time steps ($\approx$ one $ms$ in Argon units), without any signs of crystallization. A log-log plot of the mean square displacement of the A Particles for the modified system at four different temperatures (T=1.00, 0.40, 0.275, 0.25) is shown in Figure 4. At the low temperatures the ballistic- and diffusive time-regimes are very well separated, and consequently the plateau between the two regimes is very well-defined. The B- particles behaves in a similar way; but as the temperature is decreased the ratio of the diffusion coefficients D(B)/D(A) changes from 1.7 for T=1.5 to 9.7 for T=0.25. 

\begin{figure}
    \includegraphics[width=12cm]{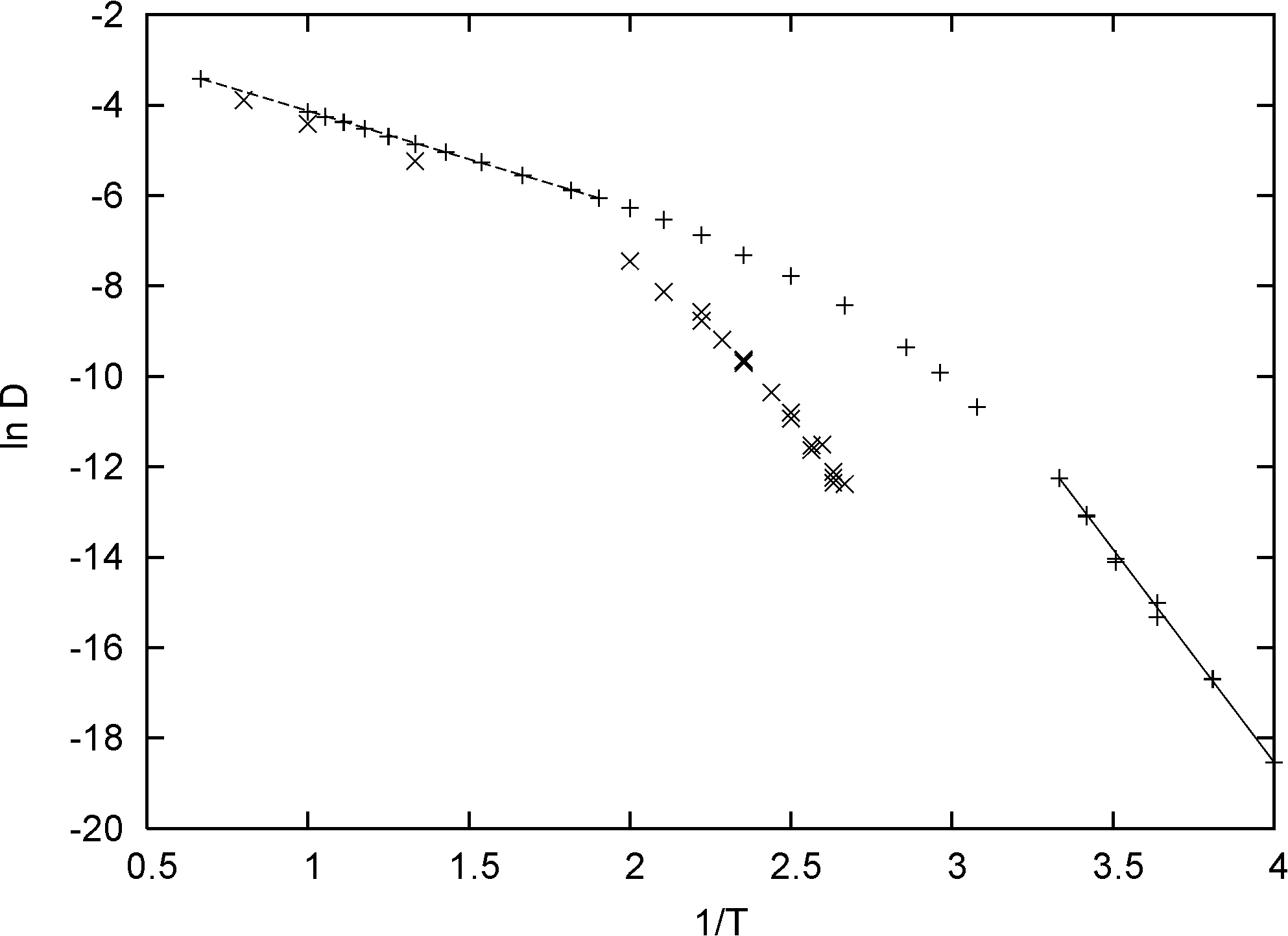}
\caption{ An Arrhenius plot, lnD(1/T), of the self diffusion constant D for the A-particles. With + is D for the modified mixture, and the points given by $\times$ is for the KA-mixture. The two straight lines show the two Arrhenius behavior. The B-particles  behave in a similar way.}
\end{figure} 

\begin{figure}
    \includegraphics[width=12cm]{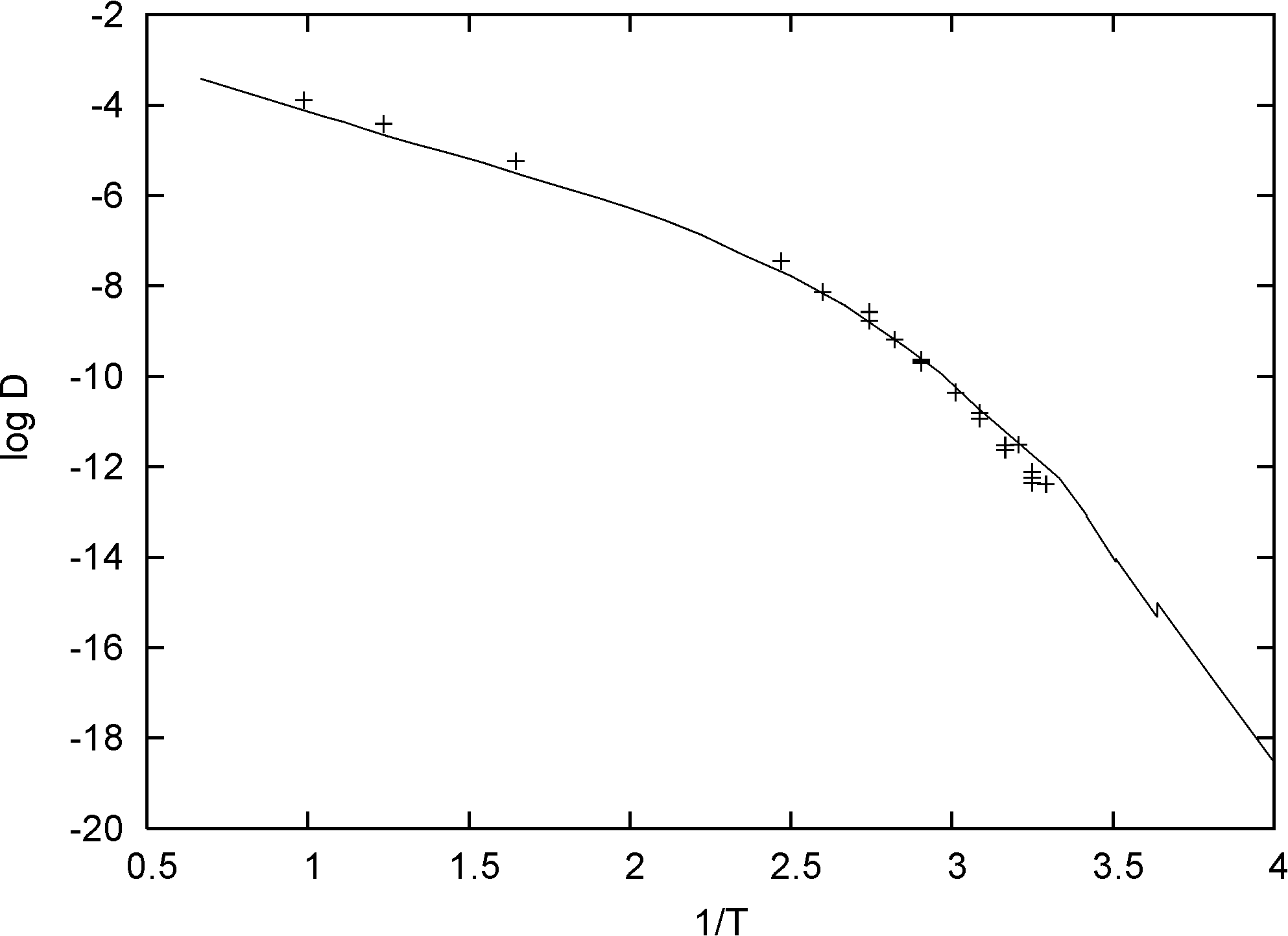}
\caption{ With full line is the ln(D)(1/T) for the modified mixture ( the data points from Fig.8 are just connected by straight lines) and with dashes are the corresponding scaled KA-data ln(D)/1/(0.81*T)}
\end{figure}

The diffusion constants (for A-particles) for the two mixtures are shown in Figure 5. Lines indicate two regions of 
(approximate) Arrhenius behavior, one at high temperatures and one at low temperatures.  A similar behavior for the KA- mixture has been reported \cite{sastry}.

It is possible to scale the  diffusion constants for the two mixtures as is demonstrated in Figure 6, where ln(D) for the KA-mixture is plotted as a function of 1/(T*0.81). A similar behavior for
"fragility invariance" for systems with the same repulsive potentials has
been observed before \cite{con}. 

The modified mixture disobeys the LB-energy rule much more than the KA-mixture. The LB-energy rule is a geometric mean of interaction energy between two
particles with  energy scaling parameters, $\epsilon_{\textrm{AA}}$ and  $\epsilon_{\textrm{BB}}$,
The two models have the same energy scaling parameters, $\epsilon_{\alpha \beta}$; but when judging the 
the disagreement with the LB-energy rule (Eq. 2) for the modified system we should substitute the minimum 
of the interaction energy for $\epsilon_{\alpha \beta}$, i.e. the right-hand side of Eq. 2 is zero, as a 
result of the "cut and shift" applied. The result is that the B-particles are almost "covalently" bounded to the A-particles, and the violation of the LB-rule for the mixing energy gives a recipe of creating fragile and stable supercooled liquid mixtures. Supercooled WCA-type mixtures with repulsive LJ-potentials have been studied recently where {\it all} interactions are WCA \cite{Evans,Chandler}. We find, however, that such systems always crystallize upon lengthy computer runs.

\end{document}